\documentclass[11pt,onecolumn,oneside]{pnas-new}

\templatetype{pnasresearcharticle} 

\title{Spontaneous formation of geysers at only one pole on Enceladus' ice shell}

\author[a,1]{Wanying Kang}
\author[a]{Glenn Flierl} 

\affil[a]{Earth, Atmospheric and Planetary Science Department, Massachusetts Institute of Technology}

\leadauthor{Kang} 

\significancestatement{Enceladus, an icy moon of Saturn, is one of a handful of places where the existence of liquid water has been confirmed. Vapor emanated out of geysers over the south pole provides a unique opportunity for us to peek inside. However, why the ice shell is much thinner over the south pole than anywhere else, including the north pole, remains a mystery. Our work aims to provide an explanation for this unexpected hemispheric asymmetry. We found that, from an infinitesimal random perturbation early in the moon's history, a significant level of asymmetry can build up over millions of years. Eventually, the ice over one of the two poles could crack open.}

\correspondingauthor{\textsuperscript{1}To whom correspondence should be addressed. E-mail: wanyingkang@g.harvard.edu}


\begin{abstract}
  The ice shell on Enceladus, an icy moon of Saturn, exhibits strong asymmetry between the northern and southern hemispheres, with all known geysers concentrated over the south pole, even though the expected pattern of tidal-rotational deformation should be symmetric between the north and south poles. Using an idealized ice evolution model, we demonstrate that this asymmetry may form spontaneously, without any noticeable a priori asymmetry (such as a giant impact or a monopole structure of geological activity), in contrast to previous studies. Infinitesimal asymmetry in the ice shell thickness due to random perturbations are found to be able to grow indefinitely, ending up significantly thinning the ice shell at one of the poles, thereby allowing fracture formation there.
  Necessary conditions to trigger this hemispheric symmetry breaking mechanism are found analytically. A rule of thumb we find is that, for Galilean and Saturnian icy moons, the ice shell can undergo hemispheric symmetry breaking only if the mean shell thickness is around 10-30~km.
\end{abstract}

\dates{This manuscript was compiled on \today}

\begin{document}

\maketitle
\thispagestyle{firststyle}
\ifthenelse{\boolean{shortarticle}}{\ifthenelse{\boolean{singlecolumn}}{\abscontentformatted}{\abscontent}}{}

Despite its small size (252 km in radius) and hence rapid heat loss, Enceladus, an icy moon of Saturn, retains a global ocean underneath its ice shell \cite{Hemingway-Iess-Tadjeddine-et-al-2018:interior, Roberts-Nimmo-2008:tidal}; geyser-like jets of water, methane, and other volatiles are shot out of the ice shell at the south pole \cite{Hansen-Esposito-Stewart-et-al-2006:enceladus, Spencer-Nimmo-2013:enceladus, Waite-Combi-Ip-et-al-2006:cassini, Waite-Glein-Perryman-et-al-2017:cassini}. These unique characteristics imply a high astrobiological potential, which has triggered interest while introducing puzzles. One of the puzzles is why all these geysers, and hence most of the heat flux, are concentrated near the south pole \cite{Howett-Spencer-Pearl-et-al-2011:high, Spencer-Howett-Verbiscer-et-al-2013:enceladus, Iess-Stevenson-Parisi-et-al-2014:gravity, Porco-Spitale-Mitchell-et-al-2007:enceladus}. This puzzle is twofold: we need to understand why the geysers tend to gather to one spot, as well as why the spot is located at the south pole, knowing that the expected pattern of tidal heating is expected to be almost perfectly symmetric between the north and south hemispheres \cite{Chen-Nimmo-2011:obliquity,Baland-Yseboodt-Van-2016:obliquity}. Previous studies have achieved the observed dichotomy by imposing an a priori anomaly in the south polar ice shell, mechanically or thermally, initially or constantly \cite{Han-Showman-2010:coupled, Han-Tobie-Showman-2012:impact, Behounkova-Tobieb-Chobletb-et-al-2012:tidally, Rozel-Besserer-Golabek-et-al-2014:self}. Despite successes, these hypotheses suffer from a common issue: the origin of the asymmetry relies on either a giant impact or a monopole structure of geological activity, followed by true polar wandering \cite{Nimmo-Pappalardo-2006:diapir, Stegman-Freeman-May-2009:origin, Tajeddine-Soderlund-Thomas-et-al-2017:true}, which requires some luck to form only one hot-spot, as observed. 

\section*{Toward an explanation without a priori asymmetry}

A concurrent work \cite{Hemingway-Rudolph-Manga-2019:cascading} proposes that the ice shell are torn apart by the overpressure induced by a secular freezing of the subsurface ocean. This occurs at poles because the ice shell is thinnest and thus weakest there; it occurs at only one of the poles because once the initial fracture forms at one pole, the overpressure would be released, preventing the same fracture from forming at the other pole. 

Our work aims to propose an alternative mechanism to explain the gathering of geysers over the south pole, which does not require any significant a priori asymmetry or secular cooling. We will show that a hemispherically symmetric ice shell is not a stable equilibrium, and that hemispheric symmetry breaking could be triggered by an infinitesimal random perturbation. Two key characteristics are necessary for a symmetry-breaking mechanism: 1) it should involve some positive feedbacks that amplify the existing inhomogeneity in the ice shell and 2) it should be able to select large-scale inhomogeneities compared to small-scale ones, so that the ice shell eventually will be dominated by large-scale topography, instead of becoming ``patchy'' and having geysers spread all over the globe.

\section*{Overview of our ice evolution model}

To understand why Enceladus ends up with a hemispherically asymmetric ice shell, we build a model to calculate how ice shell geometry would evolve. A similar problem has been looked into by \cite{Ojakangas-Stevenson-1989:thermal}, who analytically solved the equilibrium state of the ice shell geometry, where cooling induced by heat conduction balances the tidal heat generation. Here, we follow a similar path, but in addition to solving for the equilibrium state, we also look into the stability of such a state.

\begin{figure}[tbhp]
  \centering
  \includegraphics[width=0.4\textwidth]{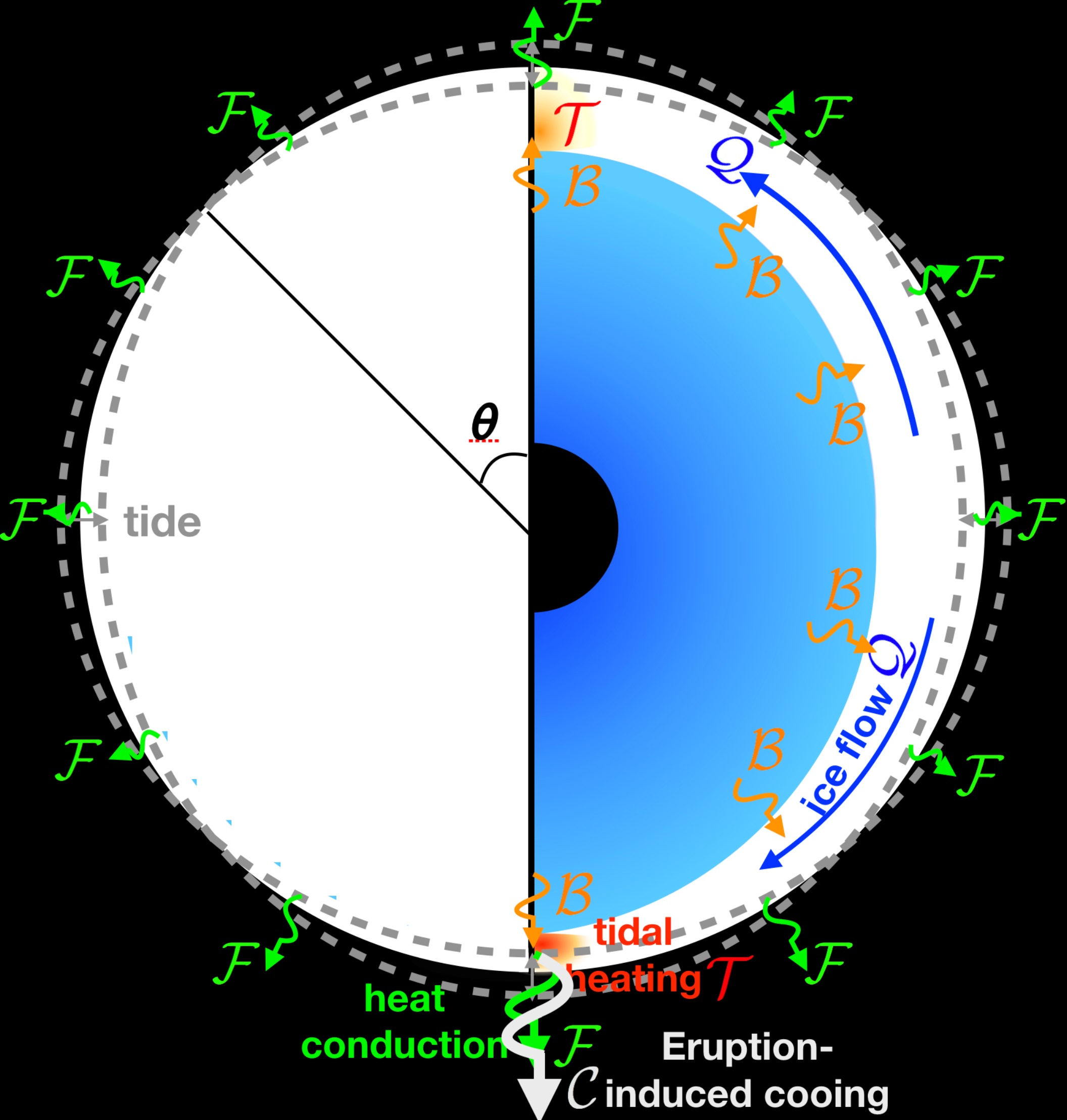}
  \caption{Schematics to demonstrate the physics processes we consider in this model. See the main text for descriptions.}
  \label{fig:schematics}
\end{figure}

As sketched in Fig.~\ref{fig:schematics}, we consider a global ice sheet whose thickness profile $H$ is simultaneously reshaped by the freezing induced by heat conduction to space $\mathcal{F}$ (outward-pointing green curly arrows), melting induced by tidal heating $\mathcal{T}$ (redish patches in the ice shell), mass redistribution by lateral ice flow $\mathcal{Q}$ (blue arrows in the ice shell), crack-induced cooling $\mathcal{C}$, and an extra heating $\mathcal{B}>0$. The governing equation of ice thickness evolution can be symbolically written as 
\begin{equation}
  \label{eq:ice-evolution}
  \frac{dH}{dt}=\frac{\mathcal{F}(H)-\mathcal{T}(H)}{L_f\rho_i} + \frac{1}{a\sin\theta}\frac{\partial}{\partial \theta} \left(\sin\theta \mathcal{Q}(H)\right)-\mathcal{B}(H)+\mathcal{C}(H).
\end{equation}
In the above equation, $H$ denotes ice thickness, $\rho_i$ denotes ice density, $L_f$ denotes fusion energy, $\theta$ denotes colatitude, and $a$ denotes the radius of Enceladus. $\mathcal{F},\ \mathcal{T},\ \mathcal{Q},\ \mathcal{B},\ \mathcal{C}$ are all functions of the ice thickness $H$. Their definitions are given in the supplementary material.

The heat conduction to space $\mathcal{F}$ is inversely proportional to $H$. The ice flow $\mathcal{Q}$ flows down-gradient, smoothing the ice shell inhomogeneities, particularly those at small scales. $\mathcal{Q}$ weakens in regions with thin ice, allowing the formation of ice ``holes''. The crack-induced cooling $\mathcal{C}$ allows extra heat loss when the ice thickness is below the crack threshold $H_{\mathrm{crack}}=8$~km (to account for the strong heat flux from the geysers on the south pole \cite{Howett-Spencer-Pearl-et-al-2011:high, Spencer-Howett-Verbiscer-et-al-2013:enceladus, Iess-Stevenson-Parisi-et-al-2014:gravity, Porco-Spitale-Mitchell-et-al-2007:enceladus}). The balancing term $\mathcal{B}$ is introduced to keep the global-mean ice thickness $H_0$ unchanged, as our focus here is to understand topography formation on the ice shell rather than the maintenance of the ice shell at a certain thickness.

Tidal heating generated in the ice shell and the core dominantly balances heat loss by conduction $\mathcal{F}$ \cite{Chen-Nimmo-Glatzmaier-2014:tidal, Matsuyama-Beuthe-Hay-et-al-2018:ocean}. As Enceladus orbits Saturn, the change of geopotential induced by orbital eccentricity deforms the ice shell and the core and generates heat. Here, we focus on the tidal dissipation in the ice shell, which is sketched as two reddish patches over the poles to demonstrate the polar-amplified heating profile, as suggested by previous studies \cite{Beuthe-2013:spatial, Beuthe-2018:enceladuss, Beuthe-2019:enceladuss, Hemingway-Mittal-2019:enceladuss}.

In a laterally varying ice shell, tidal heating $\mathcal{T}$ would be concentrated in regions where the ice shell is thinner and thus more mobile (geometric effect), just like a rubber band with a weak point would generate more heat there. The enhanced heating rate over the thin ice regions can, in turn, warm up the ice shell, making it even more mobile (rheology feedback). To make the problem analytically approachable, we account for this effect by amplifying the membrane mode heat generation by a factor of $(H/H_0)^{p_\alpha}$\footnote{Among the three tidal dissipation modes, the membrane mode is the dominant one. Physically, it corresponds to the dissipation induced by extension/compression and tangential shearing of the ice membrane \cite{Beuthe-2019:enceladuss}.}. According to the calculation in \cite{Beuthe-2019:enceladuss}, the geometric effect contributes $-1$ to $p_\alpha$, and rheology feedback contributes an additional $-0.6$ (see supplementary material for details). Besides these, a more dissipative ice shell would have an even stronger rheology feedback \cite{Beuthe-2019:enceladuss}, and ocean circulation may also contribute (see Conclusions and Discussions). Given these uncertainties, we leave $p_\alpha$ as a tunable parameter.
The schematics depict a situation where the ice shell is thinner over the south pole and thus tidal heating is amplified there. 

The idealized Maxwell rheology used here has been shown to underestimate the total tidal heat generation \cite{McCarthy-Cooper-2016:tidal,Renaud-Henning-2018:increased}. To compensate for this underestimation, a multiplicative factor, $\gamma$, is incorporated in the tidal heating formula. Since the total tidal heating in the ice shell is poorly constrained, we leave $\gamma$ as a tunable parameter. Code and data used in this study can be downloaded from \href{https://github.com/wanyingkang/icy-moon-shell-evolution.git}{here}. 

\section*{Symmetry breaking of Enceladus' ice shell}
\begin{figure*}[tbhp]
  \centering \includegraphics[width=0.75\textwidth]{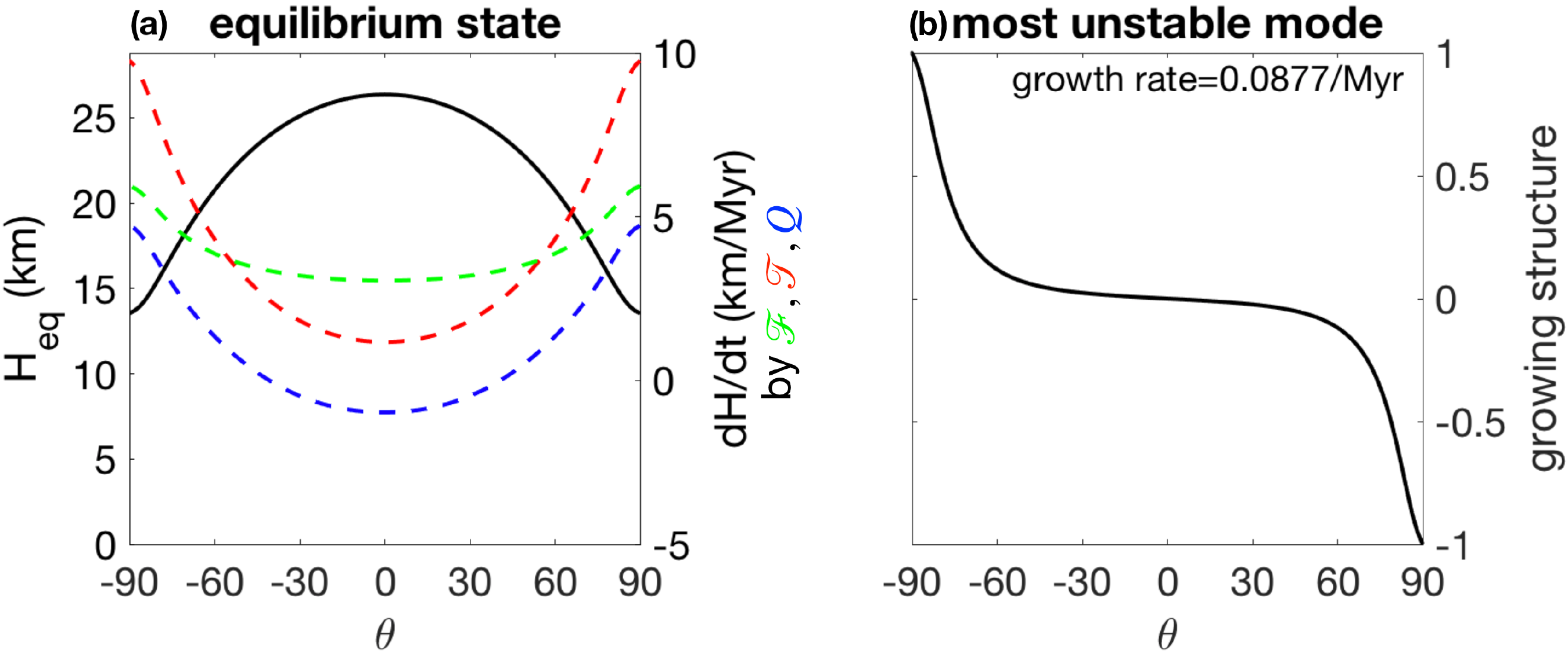}
  \caption{The equilibrium ice thickness starting from a perfect uniform ice shell and the most unstable mode based on linear analysis. Panel (a) shows $H_{eq}$ as a function of colatitude $\theta$ as a solid black curve with the y-axis on the left. The equilibrium-state melting/freezing rate induced by tidal heating $\mathcal{T}(H)/L_f\rho_i$ (red dashed), ice flow $\frac{1}{a\sin\theta}\partial_\theta \left(\sin\theta \mathcal{Q}(H)\right)$ (blue dashed), and heat loss to space $\mathcal{F}(H)/L_f\rho_i$ (green dashed) are shown on the right axis. A minus sign is multiplied to the tidal heating curve and heat loss curve to show in the same axis. Panel (b) shows the normalized structure of the most unstable mode against colatitude $\theta$ and the corresponding growth rate.}
  \label{fig:equilibrium-linear-unstable-mode}
\end{figure*}

The equilibrium ice topography $H_{eq}$ can be obtained numerically by evolving the ice thickness model from an initial value $H(\theta)=H_0$ for a long enough period of time until the tendency terms exactly compensate each other. $H_{eq}$, and the corresponding melting/freezing rate induced by $\mathcal{F},\ \mathcal{T},\ \mathcal{Q}$ are shown in Fig.~\ref{fig:equilibrium-linear-unstable-mode}a. The tidal heating $\mathcal{T}$ peaks at the two poles. Compensated by a relatively faster heat loss to space $\mathcal{F}$ in the polar regions and a poleward ice flow, the system reaches equilibrium, with thinner ice at the two poles. Here, we set $\gamma=38$, which yields a $24.8$~mW/m$^2$ tidal heating rate on global average. Since this heating rate is lower than the average heat loss rate at $\mathcal{F}=33.8$~mW/m$^2$ given by the heat conduction model, a positive constant balancing heating $\mathcal{B}$ is required to keep the global ice shell thickness unchanged. Physically, this term could correspond to dissipation in the ocean or the core \cite{Choblet-Tobie-Sotin-et-al-2017:powering, Matsuyama-Beuthe-Hay-et-al-2018:ocean, Hay-Matsuyama-2019:nonlinear}.

\begin{figure}[h!]
  \centering
  \includegraphics[width=0.4\textwidth]{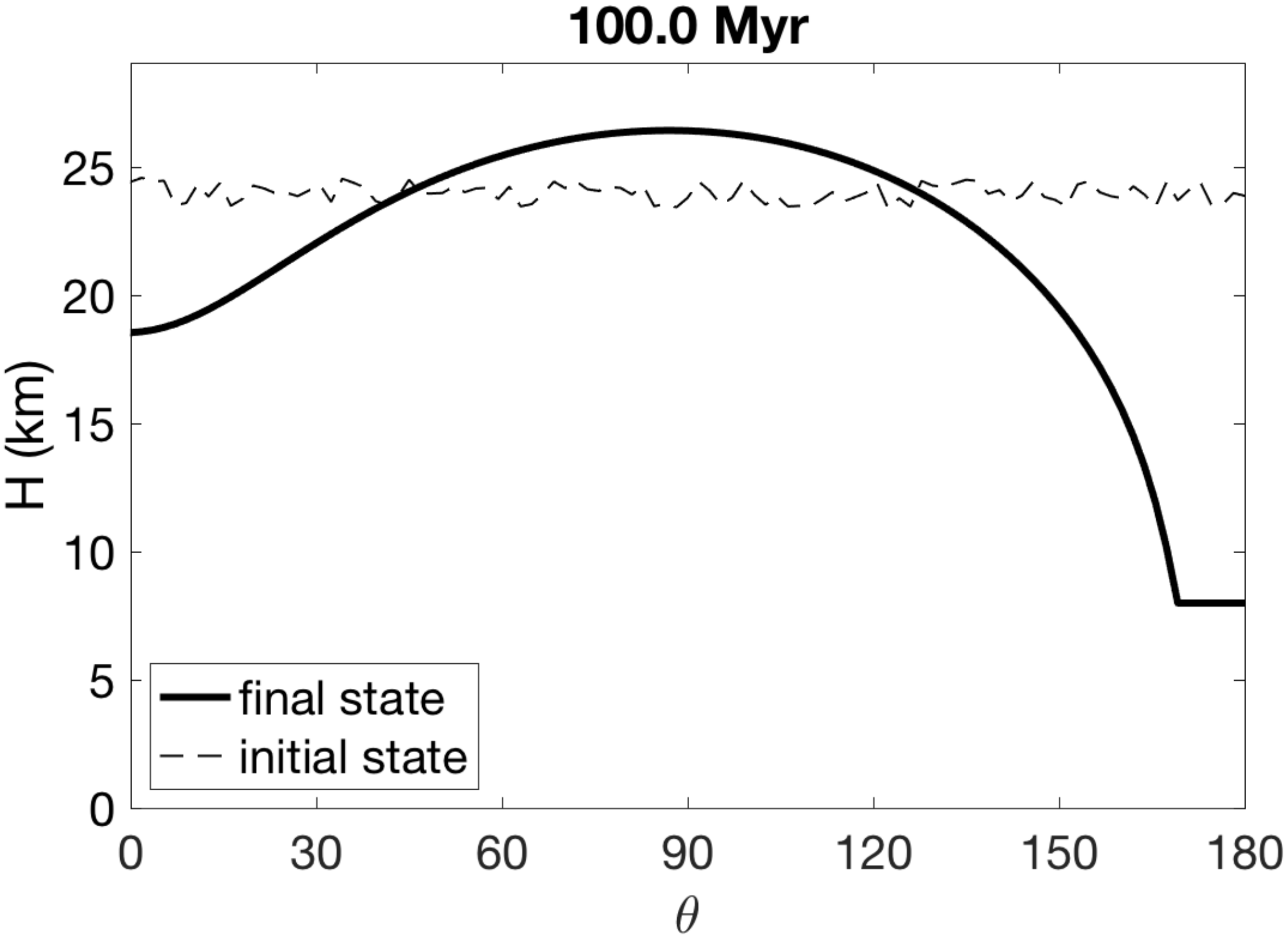}
  \caption{The final ice topography $H$ (thick solid curve) after evolving the ice thickness model (Eq.~\ref{eq:ice-evolution}) for 100~Myr. Initial condition of $H$ is shown as a thin dashed curve for comparison. The initial $h'=(H-H_0)/H_0$ at each grid point is independent and identically drawn from a uniform distribution between -0.025 and 0.025. The system would equilibrate faster with larger initial perturbations. $p_\alpha$ is set to $-1.5$.}
  \label{fig:initial-final-h}
\end{figure}

In the real world, we would not expect a perfectly uniform ice shell to begin with. If the initial ice shell thickness $H$ is slightly and randomly perturbed around the mean $H_0$, as shown by the thin dashed curve in Fig.~\ref{fig:initial-final-h}, a break in symmetry between the two hemispheres appears spontaneously. After 100 million years, the ice sheet thickness reaches a final state with a significant tilting from one pole to the other, as shown by the thick solid curve in Fig.~\ref{fig:initial-final-h}. This final state captures the main characteristics of ice-shell geometry constrained by observation \cite{Cadek-Tobie-Van-et-al-2016:enceladuss, Beuthe-2016:crustal, Hemingway-Mittal-2019:enceladuss}: the ice shell is thinner at the poles than the equator, with geysers concentrated at one of the poles.

This symmetry breaking arises from a normal mode instability induced by the concentration of tidal heating in the regions where the ice shell is already thinner than elsewhere. By linearizing the ice evolution model around the unperturbed equilibrium state $H_{eq}$, we obtain the linear tangential system (see the last section in the supplementary material for derivation). The most unstable eigenmode structure is shown in Fig.~\ref{fig:equilibrium-linear-unstable-mode}b and has a pole-to-pole tilting structure. If this structure continues to grow with time, the ice shell over one of the poles (depending on the initial condition) would get thinner and thinner, and finally a parallel set of geysers may develop through the mechanism proposed in \cite{Hemingway-Rudolph-Manga-2019:cascading}, as we see in Fig.~\ref{fig:initial-final-h}.

\section*{Discussions}
 The above calculation suggests that, without a priori asymmetry, Enceladus could naturally evolve into a state with a significant hemispheric asymmetry, and with one pole being concentrated with geysers. With the present setup ($p_\alpha=-1.5$, surface temperature $T_s=80$~K), symmetry breaking occurs when $\gamma\in [36.9,38.1]$, which corresponds to a global-mean ice dissipation rate of $23.8$-$25.1$~mW/m$^2$. Below this range, the hemispheric asymmetry of the ice shell would stabilize at a hemispheric symmetric state, with no geysers anywhere. Above this range, the topography self-amplification becomes so strong that not only the degree-1 mode but also the degree-2 mode would become unstable. The degree-2 mode could dominate over degree-1 because the polar-amplified tidal heating profile would force a degree-2 pattern in $H_{\mathrm{eq}}$, even without perturbations. As a result, geysers (ice thinner than 8~km) would form in both poles. Symmetry breaking occurs within a relatively narrow range of $\gamma$ for most of the parameters we explore here because $\gamma$ needs to be large enough for the degree-1 mode to grow while small enough for other modes (in particular degree-2 mode) to decay.

\begin{figure*}[tbhp]
  \centering
  \includegraphics[width=\textwidth]{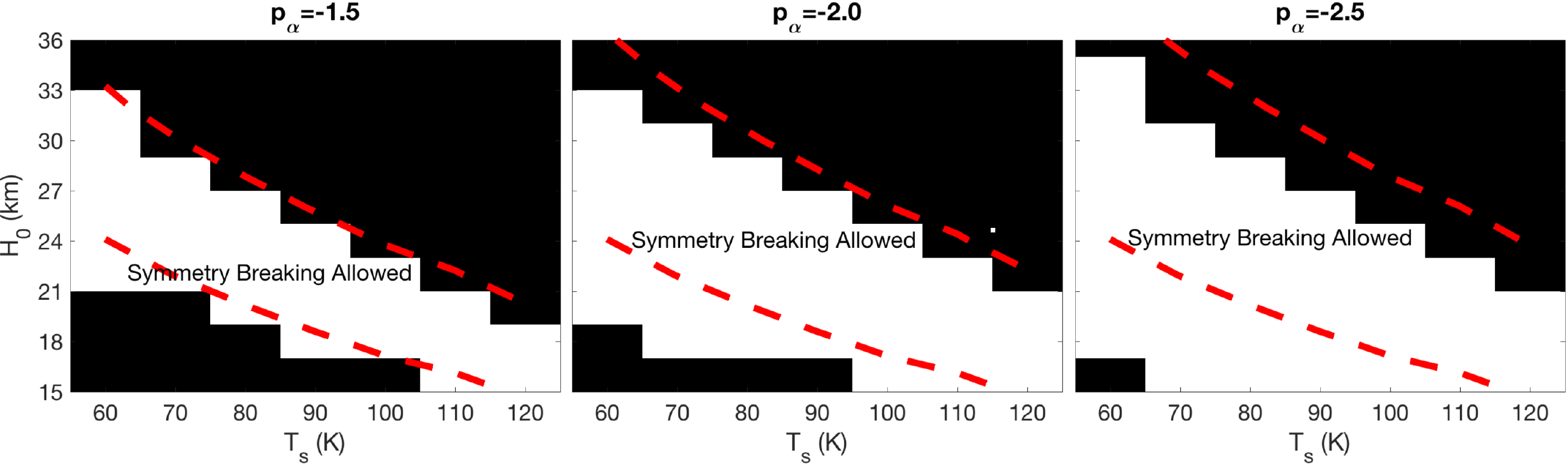}
  \caption{Diagram showing the combinations that permit symmetry breaking of the global mean ice shell thickness $H_0$ and the surface temperature $T_s$. White denotes the parameter regimes that allow us to find a $\gamma$ that can lead to symmetry breaking. Red curves show the symmetry breaking criteria given by Eq.~(\ref{eq:symmetry-breaking-criteria}) with $\beta=1, \tau_{\mathcal{F}}/\tau_{\mathcal{T}}=1$.}
  \label{fig:symmetry-break-regime}
\end{figure*}

 To show that the symmetry-breaking mechanism is not completely sensitive to the choice of poorly constrained parameters, we repeat the same calculation for various ($H_0$, $T_s$) combinations, guided by observational constraints \cite{Cadek-Tobie-Van-et-al-2016:enceladuss, Beuthe-Rivoldini-Trinh-2016:enceladuss, Hemingway-Iess-Tadjeddine-et-al-2018:interior} and for three different $p_\alpha$ values. For each $(H_0,T_s)$ combination, we search for a $\gamma$ that can lead to the hemispheric symmetry breaking, and meanwhile, guarantees that the global mean tidal heating rate wouldn't exceed the global mean heat loss to space\footnote{This leaves room for other heat sources.}. The combinations that allow us to find such a $\gamma$ are marked by the white color in Fig.~\ref{fig:symmetry-break-regime}. The three panels show results for $p_\alpha=-1.5,\ -2,\ -2.5$, respectively.

 As $|p_\alpha|$ increases, the symmetry breaking regime gradually widens, as does the range of $\gamma$ that can trigger the symmetry breaking. This is unsurprising because a greater $|p_\alpha|$ can enhance the growth of ice topography, which is one of the two necessary elements for symmetry breaking to happen.

 From Fig.~\ref{fig:symmetry-break-regime}, one can clearly see that, for any given $p_\alpha$ and $T_s$, symmetry breaking can only occur in a specific range of $H_0$. With an ice thickness beyond the upper bound (no symmetry breaking regimes are found for $H_0\leq 36$~km), symmetry breaking does not occur because the ice flow becomes too efficient\footnote{As mentioned in the model overview and shown in the supplementary material, ice flow becomes more (less) efficient when ice shell is thick (thin).} to maintain any topography. With ice thickness below the lower bound, the ice flow is too weak to maintain a crack-free $H_{\mathrm{eq}}$ in absence of perturbations. According to the stability analysis shown in the supplementary material, the symmetry breaking regime should satisfy the following criteria:
 \begin{eqnarray}
   \frac{\beta(\tau_{\mathcal{F}}/\tau_{\mathcal{T}})}{4}<\frac{\tau_{\mathcal{F}}}{\tau_{\mathcal{Q}}}<-p_\alpha(\tau_{\mathcal{F}}/\tau_{\mathcal{T}})(\beta/2+1)-1 \label{eq:symmetry-breaking-criteria}  
 \end{eqnarray}
 Here, $\tau_{\mathcal{F}},\ \tau_{\mathcal{T}},\ \tau_{\mathcal{Q}}$ are the timescales for heat conduction, tidal heating, and ice flow that significantly change the planetary-scale ice topography, and $\beta$ is the percentage variation of $\mathcal{T}$ about the mean across different latitudes (see supplementary material for definitions).
 As shown by red dashed curves in Fig.~\ref{fig:symmetry-break-regime}, Eq.~\ref{eq:symmetry-breaking-criteria} predicts the location of the symmetry breaking regime in the ($T_s$,$H_0$) parameter space well, especially for $p_\alpha=-1.5$. For greater $|p_\alpha|$, $\tau_{\mathcal{F}}/\tau_{\mathcal{T}}$ becomes significantly smaller than 1 when symmetry breaking occurs; therefore, the actual symmetry breaking regime is more on the low side of $H_0$, compared to the prediction given by Eq.~\ref{eq:symmetry-breaking-criteria}.

For such a symmetry breaking regime to exist, the upper bound of Eq.~(\ref{eq:symmetry-breaking-criteria}) has to be larger than the lower bound, which leads to an additional necessary condition for ice shell symmetry breaking,
\begin{equation} \frac{\tau_{\mathcal{F}}}{\tau_{\mathcal{T}}}>\frac{1}{-p_\alpha(\beta/2+1)-\beta/4}.
  \label{eq:symmetry-breaking-criteria2}
\end{equation}
This sets a lower bound for the percentage of heat loss that is compensated by the tidal heating in the ice shell.
For the example, as we show in the results, this condition would be satisfied if the ice dissipation rate is enhanced by a $\gamma$ factor greater than 14.8, which could be possible given that the Maxwell body rheology used here could significantly underestimate the dissipation rate \cite{McCarthy-Cooper-2016:tidal,Renaud-Henning-2018:increased}.


The ice shell symmetry breaking criteria (Eq.~\ref{eq:symmetry-breaking-criteria} and Eq.~\ref{eq:symmetry-breaking-criteria2}) can be applied to other icy moons, if their orbital parameters and physical characteristics are known. 
For icy moons around Jupiter and Saturn whose ice shell is a major heat producer (i.e., Eq.~\ref{eq:symmetry-breaking-criteria2} is satisfied), symmetry breaking would occur at roughly the same $H_0$, regardless of other parameters. Substituting the definitions of $\tau_{\mathcal{F}}$ and $\tau_{\mathcal{Q}}$ into Eq.~(\ref{eq:symmetry-breaking-criteria}) and noticing that both the lower and upper bound of that inequality are of $O(1)$, we get
\begin{equation} \frac{\tau_{\mathcal{F}}}{\tau_{\mathcal{Q}}}\propto \frac{H_0^5}{(a^2/g)(\overline{\eta}\log\left(T_m/T_s\right))}\sim O(10^{-8}),\label{eq:tauF-tauQ}
\end{equation}
where $g$ and $a$ are the surface gravity and radius of Enceladus, and $\overline{\eta}$ is a weighted average viscosity across the ice shell, whose definition is given in the supplementary material. The $10^{-8}$ on the right comes from a combination of physical constants, including fusion energy, thermal conductivity, and density of the ice. The fact that $\left.\tau_{\mathcal{F}} \right/\tau_{\mathcal{Q}}$ is proportional to the fifth power of $H_0$ implies the symmetry breaking regime is always centered around $H_0\sim 10$-$30$~km. Even if the radius of the moon is enlarged by 10 times (while keeping the bulk density unchanged), $a^2/g$ would increase by a factor of 10, and the symmetry breaking permitting $H_0$ would only change by a factor of $(10)^{1/5}\sim 1.6$. The Galilean satellites are warmer than Saturnian satellites by about 30K, and the associated impact on $H_0$ is no larger than a factor of $2$ as well. This leads to an even simpler empirical rule for the mechanism proposed here: ice shell symmetry breaking would only occur on Galilean and Saturnian icy moons whose ice shell thickness is around 10-30~km.

Our calculation assumes the ice shell is conductive. However, besides Enceladus, the only place inside our solar system that satisfies this assumption is Europa; other icy moons (such as Titan, Ganymede, Callisto) have an ice shell that is too thick \cite{Schenk-2002:thickness, Kuskov-Kronrod-2005:models, Hemingway-Nimmo-Zebker-et-al-2013:rigid, Spohn-2010:treatise} for ice convection to be triggered \cite{McKinnon-2006:convection}. Repeating the same calculation for Europa, we found symmetry breaking could only occur when the ice shell thickness is 20-26~km, assuming the mean surface temperature is around 100~K \cite{McFadden-Weissman-Johnson-2007:encyclopedia} and $p_\alpha=-1.5$. Based on magnetic conduction, Europa's ice shell thickness is constrained to be below 15~km \cite{Hand-Chyba-2007:empirical}, indicating that Europa could have narrowly missed the symmetry-breaking regime. However, the estimation of Europa's ice shell thickness is not yet conclusive \cite{Tobie-Choblet-Sotin-2003:tidally, Schenk-2002:thickness, Nimmo-Giese-Pappalardo-2003:estimates}, so more studies are needed to improve the thickness estimation and the symmetry breaking criteria.

  To make the problem analytically approachable, we made many simplifications and assumptions, which could have an impact on the symmetry breaking regime. For example, when considering the tidal heating generated in an inhomogeneous ice shell, we multiply $\mathcal{T}_0^{\mathrm{mem}}$ by a factor of $(H/H_0)^{p_\alpha}$ rather than solve the thin shell ODE with variable coefficients. As a result, we ignore the fact that small-scale perturbations would be suppressed by the greater bending rigidity associated with it. Taking this into consideration would lead to a stronger scale selectivity. In addition, the slow circulation in the slush zone at the ocean-ice interface can also help damp the small-scale topographies, widening the symmetry breaking regime. We also ignore the decrease of surface temperature toward the poles, as the role played by this factor is less clear and will be explored in future work.
 
Other ignored mechanisms, such as ocean circulation, surface snow cover, dynamics of ice convection, and non-Newtonian ice rheology, could also affect the parameter regime where this mechanism works and the rate at which hemispheric asymmetry grows.
Particularly, tidal heating from the core is found to be necessary to counterbalance the heat conduction to space, as heating generated in the ice shell alone is not enough, given our current understanding of ice dynamics \cite{Beuthe-2019:enceladuss, Hemingway-Mittal-2019:enceladuss}. The heating generated in the core is carried upward by eddies, convection plumes, and large-scale circulation in the ocean. It is possible that some of these processes could respond to changes in ice topography and form a feedback loop, as suggested by \cite{Ashkenazy-Sayag-Tziperman-2018:dynamics}. We speculate that freezing (melting) at the ocean-ice interface may be able to reduce (increase) the local salinity and drive an upwelling (sinking) motion there, which in turn enhances (reduces) the upward heat transport from the core and leads to further freezing (melting). Feedback like this can be accounted for in our framework by a greater $p_\alpha$ factor. Processes that are not directly affected by the ice topography are represented by the balancing term $\mathcal{B}$. Our work assumes a globally uniform profile of $\mathcal{B}$, but it is highly possible that $\mathcal{B}$ is concentrated in a few hot spots \cite{Choblet-Tobie-Sotin-et-al-2017:powering}. If that is true, the equilibrium state $h_{\mathrm{eq}}'$ would change, which would in turn affect the growing mode structure and its growth rate. Even a tiny hemispheric asymmetry in $\mathcal{B}$ would help determine the direction of symmetry breaking, and it could significantly affect the parameter regime where symmetry breaking could happen.

\section*{Concluding remarks}

Our work takes an idealized framework to demonstrate the feasibility of spontaneous hemispheric symmetry breaking.
Over billions of years, Enceladus could have gone through large eccentricity variations \cite{Meyer-Wisdom-2007:tidal, Yoder-1981:tidal, Ojakangas-Stevenson-1986:episodic}, leading to secular melting/freezing over time. As the ice shell thickness varies, it seems unavoidable that Enceladus would pass through the symmetry breaking regime shown in Fig.~\ref{fig:symmetry-break-regime}. When Enceladus gets into the symmetry breaking regime, hemispheric asymmetry would accumulate; when it gets out, hemispheric asymmetry would gradually decay. Before the historical orbital record is obtained, understanding the stochastic dynamics of this system would shed light on how much hemispheric asymmetry of Enceladus can be developed through the mechanism proposed here.

\showmatmethods{} 

\acknow{Most of this work was carried out during the 2019 Geophysical Fluid Dynamics Summer School at Woods Hole. We thank the GFD faculty for organizing the program and the Woods Hole Oceanographic Institution for hosting it. During the review process, the two anonymous reviewers gave us extremely helpful comments, which we deeply appreciate. We also thank Prof. Geoff Vasil and Prof. Ming Cai for helpful discussions.}

\showacknow{} 

\bibliography{export}

\end{document}


\baselineskip24pt
\maketitle

\subsection*{Overview of the ice shell evolution model}
The evolution of the ice shell thickness is determined by the melting induced by the tidal heating $\mathcal{T}$, the down-gradient ice flow $\mathcal{Q}$, the heat loss to space by conduction $\mathcal{F}$, the crack-induced cooling $\mathcal{C}$ in places ice is thin enough, and an extra heating $\mathcal{B}>0$ to stabilize the global-mean ice thickness $H_0$. 
\begin{equation}
  \label{eq:ice-evolution}
  \frac{dh'}{dt}=\frac{\mathcal{F}(h')-\mathcal{T}(h')}{L_f\rho_iH_0 } + \frac{1}{a\sin\theta}\partial_\theta \left(\sin\theta \mathcal{Q}(h')/H_0\right)-\mathcal{B}/H_0+\mathcal{C}/H_0.
\end{equation}
$H$ denotes the ice shell thickness, and $h'$ is the normalized deviation from the global mean $H_0$ ($H=H_0(1+h')$). $\theta$ is the colatitude, and $a$ is radius of the moon. $L_f$ is the latent heat of ice, and $\rho_i$ is the ice density. Physical quantities are summarized in Table.~\ref{tab:parameters}. This is the same as Eq.~1 in the main text, except every term is divided by $H_0$, for convenience of the linear stability analysis (last section in this document).

The heat loss to space $\mathcal{F}$, ice flow $\mathcal{Q}$ and tidal heating $\mathcal{T}$ depend on the ice topography $h'$ through
\begin{eqnarray}
  \mathcal{F}&=&\mathcal{F}_0 (1+h')^{-1}\label{eq:F}\\
  \mathcal{Q}&=& \mathcal{Q}_0 (1+h')^3 H_0 \partial_\theta h'/a \label{eq:Q}\\
  \mathcal{T}&=&\mathcal{T}_0^{\mathrm{mem}}(\theta)(1+h')^{p_\alpha}+\mathcal{T}_0^{\mathrm{bend}} (\theta)+\mathcal{T}_0^{\mathrm{mix}} (\theta)\label{eq:H}
\end{eqnarray}

Here, $\mathcal{F}_0$ and $\mathcal{Q}_0$ are constants and $\mathcal{T}_0^{\mathrm{mem},\mathrm{mix},\mathrm{bend}}$ are meridional profiles. All these constants and profiles can be calculated given the ice properties. The key of each model is briefly summarized below. For more details, interested readers are pointed to the three methods sections following this one. 

  The heat loss to space $\mathcal{F}$ (Eq.\ref{eq:F}) is solved from a 1D heat diffusion equation, with two fixed boundary conditions: the temperature is $80~$K at the top and $273~$K at the ocean-ice interface. The heat flux $\mathcal{F}$ is inversely proportional to the ice thickness, consistent with the intuition that thicker ice preserves heat better. 

  We only consider the tidal heating contributed from the dissipation in the ice shell, because the ocean flow and the associated ocean dissipation can be strongly reduced by the crustal constraint imposed by the solid ice shell \cite{Beuthe-2016:crustal,Travis-Schubert-2015:keeping}. We follow \cite{Beuthe-2013:spatial,Beuthe-2019:enceladuss} to calculate the tidal dissipation in the ice shell $\mathcal{T}$ (Eq.\ref{eq:H}). It consists of three components: membrane mode $\mathcal{T}^{\mathrm{mem}}$ due to the extension/compression and tangential shearing of the ice membrane, mixed mode $\mathcal{T}^{mix}$ due to vertical shifting and bending mode $\mathcal{T}^{bend}$ induced by the vertical variation of compression/stretching. Among the three components, the membrane mode dominates.

  Regions with relatively thinner ice shell tend to produce more heat due to the higher extensibility there. This effect is taken into account by the factor $(H/H_0)^{p_\alpha}$ in Eq.~\ref{eq:H}. 

  The idealized Maxwell rheology used here has been shown to underestimate the total tidal heat generation \cite{McCarthy-Cooper-2016:tidal,Renaud-Henning-2018:increased}. To compensate this underestimation, a multiplicative factor, $\gamma$, is incorporated in the tidal heating formula. Since the total tidal heating in the ice shell is poorly constrained, we leave $\gamma$ as a tunable parameter.

  When topography forms in the ice shell, ice flow $\mathcal{Q}$ (Eq.\ref{eq:Q}) would homogenize it. We follow \cite{Ashkenazy-Sayag-Tziperman-2018:dynamics} using an upside down land ice sheet model to calculate the ice flow. Driven by the pressure gradient induced by the spatial variation of the ice top surface, ice flows downgradient following second order diffusion. At the top, the ice flow speed is set to zero, as the upper part of the ice shell is cold and rigid and hence hard to move. At the bottom, the vertical shear of the ice flow speed is set to zero, as required by the zero tangential stress there.
  
  
  Finally, we consider the crack-induced heat loss $\mathcal{C}$ and a balance term $\mathcal{B}$. $\mathcal{C}$ means to capture the strong heat flux around the ``tiger stripes'' \cite{Howett-Spencer-Pearl-et-al-2011:high, Spencer-Howett-Verbiscer-et-al-2013:enceladus, Iess-Stevenson-Parisi-et-al-2014:gravity, Porco-Spitale-Mitchell-et-al-2007:enceladus}. It would be triggered wherever ice thickness drops below a threshold $H_{\mathrm{crack}}=8$~km, and then any further melting in that spot would be prevented. Numerically, $\mathcal{C}$ protects the model from exhausting ice and blowing up. Since our focus is to understand topography formation on the ice shell rather than the maintenance of an ice shell at a certain thickness, we introduce $\mathcal{B}$ to stabilize the mean ice thickness at $H_0$. Physically, this term corresponds to the heat flux from below that will not vary with ice topography. We here assume $\mathcal{B}$ to be uniform over the globe, since we do not have a direct measurement.

\begin{table}[hptb!]
  \centering
  \begin{tabular}{lll}
    \hline
    Symbol & Name & Definition/Value\\
    \hline
    \multicolumn{3}{c}{Enceladus parameter}\\
    \hline
    $a$ & radius of Enceladus & 252~km\\
    $\epsilon$ & ice shell aspect ratio & $H_0/a$\\
    $e$ & orbital eccentricity & 0.0047\\
    $\omega$ & rotation rate & 5.307$\times$10$^{-5}$~s$^{-1}$\\
    $g$ & surface gravity & 0.113~m/s$^2$\\
    $T_s$ & surface temperature & 80~K\footnote{XX}\\
    $H_0$ & global averaged ice shell thickness & 24~km \cite{Beuthe-Rivoldini-Trinh-2016:enceladuss}\\
    $H_{\mathrm{crack}}$ & thickness threshold for crack to occur & 8~km \\
    \hline
    \multicolumn{3}{c}{Physical constants}\\
    \hline
    $\rho_w$ & density of ocean & 1050~kg/m$^3$ \\
    $\rho_i$ & density of ice & 950~kg/m$^3$ \\
    $\mu_e$ & elastic shear modulus of ice & 3.5~GPa \cite{Helgerud-Waite-Kirby-et-al-2009:elastic}\\
    $\nu$ & Poisson's ratio of ice & 0.33 \cite{Helgerud-Waite-Kirby-et-al-2009:elastic}\\
    $\kappa_0$ & conductivity coeff. of ice & 651~W/m \cite{Petrenko-Whitworth-1999:physics}\\
    $L_f$ & fusion energy of ice & 334000~J/kg\\
    $E_a$ & activation energy for diffusion creep & 59.4~kJ/mol \cite{Goldsby-Kohlstedt-2001:superplastic}\\
    $T_m$ & melting temperature & 273.15~K\\
    $\eta_{melt}$ & ice viscosity at $T_m$ & 10$^{14}$~Pa$\cdot$s\footnote{The ice viscosity at the melting temperature, $\eta_{\mathrm{melt}}$, has a large uncertainty and, as the shell gets thicker than a few kilometers, the melting point viscosity may decrease from $10^{16}$ to $10^{13}$ Pa$\cdot$s.}\\
    $R_g$ & the gas constant.&  8.31~J/K/mol\\
    \hline
    \multicolumn{3}{c}{Ice shell properties}\\
    \hline
    $\eta$ & viscosity of ice & $\eta_{\mathrm{melt}} \exp \left[\frac{E_{a}}{R_{g} T_{m}}\left(\frac{T_{m}}{T}-1\right)\right]$\\
    $mu$ & complex shear modulus of ice & $\frac{\mu_e}{1+\mu_e/(i\omega \eta)}$\\
    $\mu_p$ & p$^{th}$ moment of $\mu$ & $\frac{1}{H^{p+1}}\int_0^H \mu (H-z)^p~dz$\\
    $\mu_{\mathrm {inv}}$ & invariant 2$^{nd}$ moment of $\mu$ & $\mu_2 -\mu_1^2/\mu_0$\\
    $\alpha_0$ & extensibility of a $H_0$ thick ice shell & $\left(2(1+\nu) \mu_{0} H_0\right)^{-1}$\\
    $D_0$ & bending rigidity of a $H_0$ thick ice shell & $2 \mu_{\mathrm {inv }} H_0^{3} /(1-\nu)$\\
    $\chi_0$ & nondimensional ice coefficient & $\left.\left(\mu_{0}+\epsilon \mu_{1}\right) \right/\left(\mu_{0}+2 \epsilon \mu_{1}+\epsilon^2\mu_{2}\right)$\\
    & & $\approx 1-\epsilon\frac{\mu_1}{\mu_0}\approx 1+i\mathrm{Im}\left(\epsilon\frac{\mu_1}{\mu_0}\right)$\\
    \hline
  \end{tabular}
  \caption{Parameter definitions. }
  \label{tab:parameters}
\end{table}

\subsection*{Heat conduction model}
Following \cite{Beuthe-2018:enceladuss}, we assume that tidal heating is concentrated at the ice-ocean interface, and that the heat conduction within the ice shell is efficient enough so that the ice shell is always in thermal equilibrium. We can then write down the heat conduction equation
\begin{eqnarray}
  \label{eq:heat-condution}
  \frac{\partial}{\partial z}\left(\kappa \frac{\partial T}{\partial z}\right)=C_p\rho_i \frac{\partial T}{\partial t}=0.
\end{eqnarray}
$\kappa$ is the heat conductivity of ice. It varies inversely proportionally to temperature \cite{Slack-1980:thermal,Petrenko-Whitworth-1999:physics}.
\begin{eqnarray}
  \label{eq:kappa}
  \kappa=\kappa_0/T
\end{eqnarray}

Solving Eq.~\ref{eq:heat-condution} under the boundary conditions $T(0)=T_m,\ T(H)=T_s$ yields the equilibrium temperature profile,
\begin{eqnarray}
  \label{eq:temperature-profile}
  T(z)=T_{m}^{\left(H-z\right) /H} T_{s}^{z /H},
\end{eqnarray}
where $z$ is the distance above the ocean-ice interface.
We know that the temperature just above the ice-ocean interface should be close to the melting point, i.e., $T_m\approx 273$~K, and we set the temperature at the upper surface $T_s$ to $80$~K by default.

The default $T_s$ value we use here is higher than the radiative equilibrium temperature, $59$~K. This choice is to account for the insulating effect of a likely high-porosity layer on the top \cite{Travis-Schubert-2015:keeping}. In fact, the highly relaxed craters at the surface are more consistent with a relatively warmer surface temperature \cite{Bland-Singer-McKinnon-et-al-2012:enceladus}. We also do a sensitivity test to $T_s$, and the results are summarized in the Fig.~4 in the main text.

The upward heat flux at the interface is proportional to the vertical temperature gradient there,
\begin{eqnarray}
  \label{eq:interface-heat-flux-parameterization}
  \mathcal{F}=\frac{\kappa_0}{H}\log\left(\frac{T_m}{T_s}\right)\equiv \mathcal{F}_0(1+h')^{-1},
\end{eqnarray}
where $h'=(H-H_0)/H_0$ is the normalized deviation from the mean ice thickness $H_0$, $\mathcal{F}_0=\kappa_0\log\left(T_m/T_s\right)/H_0$. 
The time scale for heat conduction to adjust ice topography can be written as
\begin{equation}
  \label{eq:tau-F}
  \tau_{\mathcal{F}}\equiv \left(\frac{d(H/H_0)}{dt}\right)_{\mathcal{F}}^{-1}\sim \frac{L_f\rho_iH_0}{\mathcal{F_0}}=\frac{L_f\rho_iH_0^2}{\kappa_0\log\left(T_m/T_s\right)}
\end{equation}
The subscript $\mathcal{F}$ denotes the tendency induced by heat conduction.

\subsection*{Ice flow model}
 We use the upside-down thin ice model by \cite{Ashkenazy-Sayag-Tziperman-2018:dynamics} to calculate the vertically-integrated downgradient ice flow $\mathcal{Q}$. This model use non-slip boundary conditions at the top and free-slip boundary conditions at the bottom. The non-slip upper boundary condition seems to contradict the fact that atmosphere cannot sustain tangential stress, so we will first justify this choice.

Assuming hydrostatic balance and zero tangential stress within the ocean \cite{Iess-Stevenson-Parisi-et-al-2014:gravity,McKinnon-2015:effect,Cadek-Tobie-Van-et-al-2016:enceladuss}, we can express the ice top topography ``$s$'' using the variation of ice thickness $h=H-H_0$,
\begin{eqnarray}
  \label{eq:ice-surface-s}
  s= \frac{\rho_w-\rho_i}{\rho_w}h.
\end{eqnarray}
 To get the above formula, we ignore the vertical variation of the gravity acceleration rate $g$ and the flexural support by the ice shell, which should be accounted in a more accurate calculation \cite{Hemingway-Mittal-2019:enceladuss}. The tilted ice top exerts pressure gradient force to the ice shell below. The balance of forces requires
\begin{eqnarray}
  \label{eq:ice-stress-balance}
  \frac{1}{2}\partial_z\eta\partial_z V(\theta,z)+\frac{1}{a^2\sin\theta}\partial_\theta\eta\sin\theta \partial_\theta V(\theta,z)=\frac{\rho_ig}{a} \partial_\theta s,
\end{eqnarray}
where the ice flow speed $V(\theta,z)$ at latitude $\theta$ and height $z$.

At the top and bottom of the ice sheet, the atmosphere and ocean cannot exert tangential stress onto the ice shell (we ignored the form drag induced by the interface topography),
\begin{eqnarray}
  \label{eq:boundary-condition}
  \partial_z V(\theta,0)=\partial_z V(\theta,H)=0.
\end{eqnarray}
Here, $\eta$ is the ice viscosity. It decays exponentially with the ice temperature \cite{Schoof-Hewitt-2013:ice}, as defined in Table.~\ref{tab:parameters}. 
Substituting the equilibrium temperature profile (Eq.\ref{eq:temperature-profile}) leads to a viscosity profile that rapidly decays with depth.

Without interacting with each other, the warm and soft ice close to the interface will flow orders of magnitudes faster than the cold and rigid ice at the top, under the same amount of pressure gradient force induced by the tilting of the ice top. That is to say change of the ice flow $V(\theta,z)$ in the vertical direction is of the same magnitude as that in the horizontal direction. Since the horizontal scale of the ice flow is usually much greater than the vertical scale of the whole ice shell, the forcing due to vertical shear stress (the first term in Eq.~\ref{eq:ice-stress-balance}) will dominate that induced by the horizontal compression and extension (the second term in Eq.~\ref{eq:ice-stress-balance}). 
However, according to the boundary conditions (Eq.~\ref{eq:boundary-condition}), the vertical integral of the first term must vanish. Through this vertical shear term, the lower part of the ice shell gets connected with the upper part: it attempts to decelerate the fast ice flow in the lower part of the ice shell, while accelerating the slow ice flow in the upper ice shell. In the upper ice shell, both the pressure gradient force and the vertical shear stress from below tend to accelerate the ice flow, and they are counterbalanced by the horizontal extensional stress.

Since most of the ice flow is contributed by the soft ice layer at the bottom, our ice flow model only represents the soft ice layer, and completely ignores the slow ice flow in the upper part of the ice shell. To represent the friction provided by the motionless rigid ice above, at the top of the ice shell, we can therefore adopt a no-slip boundary condition at the top as in \cite{Ashkenazy-Sayag-Tziperman-2018:dynamics},
\begin{eqnarray}
  \label{eq:boundary-condition-noslip}
  V(\theta,H)=0.
\end{eqnarray}

Dropping the second term in Eq.~\ref{eq:ice-stress-balance} and integrating vertically, we get
\begin{eqnarray}
  \label{eq:ice-flow-thin}
  \frac{1}{2}\eta \partial_{z}V(\theta,z)=\rho_i gz\partial_\theta s/a.
\end{eqnarray}

Taking vertical integration again and substituting in the vertical profile of viscosity ($\eta$ in Table~\ref{tab:parameters}) yields
\begin{eqnarray}
  \label{eq:ice-speed}
V(\theta,z)=\frac{2(\rho_w-\rho_i)g}{\eta_{\mathrm{melt}}(\rho_w/\rho_i)\log^2\left(T_m/T_s\right)}H^2\partial_\theta h/a \int_{T_s}^{T(z)}\exp\left[-\frac{E_{a}}{R_{g} T_{m}}\left(\frac{T_{m}}{T'}-1\right)\right]\log(T')~\frac{dT'}{T'},
\end{eqnarray}
where $T'$ is an integral variable.

Finally, we take another vertical integration of $V$ from the bottom to the top of the ice shell, to get the column-integrated ice flow $Q$,
\begin{eqnarray}
  \mathcal{Q}(\theta)&\equiv&\int_0^{H}V(z)~dz= \mathcal{Q}_0\frac{H^3}{H_0^3}(\partial_\theta h/a)=\mathcal{Q}_0H_0(1+h')^3(\partial_\theta h'/a)  \label{eq:ice-flow}\\ \mathcal{Q}_0&=&\frac{2(\rho_w-\rho_i)g\zeta^3}{(\rho_w/\rho_i)\eta_{\mathrm{melt}}}.\label{eq:Q0}
\end{eqnarray}
Ice viscosity $\eta$ varies with temperature following a reciprocal exponential function (Table~\ref{tab:parameters}). We introduce $\zeta$, the effective e-fold thickness for $\eta$, to measure the thickness of mobile layer in the ice shell, following \cite{Stevenson-2000:limits}. Its definition is given below.
\begin{equation}
  \label{eq:eq:delta}
  \zeta=\frac{ H_0}{2}\left\{\int_{T_s}^{T_m}\int_{T_s}^{T(z)}\exp\left[-\frac{E_{a}}{R_{g} T_{m}}\left(\frac{T_{m}}{T'}-1\right)\right]~d\left(\frac{\ln^2 T'}{\ln^2\left(T_m/T_s\right)}\right)~d\left(\frac{\ln T}{\ln\left(T_m/T_s\right)}\right)\right\}^{1/3}.
\end{equation}

The time scale for ice flux to change ice topography can be estimated as below
\begin{equation}
   \label{eq:tau-Q}
   \tau_{\mathcal{Q}}\equiv \left(\frac{d(H/H_0)}{dt}\right)^{-1}_Q\sim \left(\mathcal{Q}_0\nabla^2(H/H_0)\right)^{-1} \sim \frac{a^2\eta_{\mathrm{melt}}}{2\zeta^3g(\rho_w-\rho_i)}\sim \frac{3}{8\pi G\rho_b(\rho_w-\rho_i)}\frac{a\bar{\eta}}{H_0^3}.
 \end{equation}
 The subscript $\mathcal{Q}$ in the second equality denotes the tendency induced by ice flow. In the above formula, we assume a global-scale ice shell topography (degree-1). The relax time scale for a degree-l mode structure should be $l^2\tau_{\mathcal{Q}}$. In the last step, we assume a fixed planetary density, which leads to $g\sim 4\pi G\rho_ba/3$, where $G$ is the gravity constant and $\rho_b$ is the bulk density of the planet. $\bar{\eta}$ is the average ice viscosity defined as
 \begin{equation}
   \label{eq:etabar}
   \bar{\eta}=\eta_{\mathrm{melt}}\left/\left\{\int_{T_s}^{T_m}\int_{T_s}^{T(z)}\exp\left[-\frac{E_{a}}{R_{g} T_{m}}\left(\frac{T_{m}}{T'}-1\right)\right]~d\left(\frac{\ln^2 T'}{\ln^2\left(T_m/T_s\right)}\right)~d\left(\frac{\ln T}{\ln\left(T_m/T_s\right)}\right)\right\}\right.
 \end{equation}

The ice flow is proportional to the slope of the ice top surface $(\partial_\theta h/a)$, and the cubic power of the ice depth $H$. Applying non-Newtonian ice rheology, such as Glen's law, would make the ice flow be proportional to the fifth power of $H$ and the third power of $\partial_\theta h$. The latter would give rise to a much stronger scale selectivity: the suppression to $l_t=10$ topography in Newtonian rheology will be experienced by $l_t=(10)^{1/3}\sim 2$ in non-Newtonian rheology.

\subsection*{Tidal heating model}
We use the tidal heating model by \cite{Beuthe-2019:enceladuss} to calculate the heating profile generated in an inhomogeneous ice shell. We here briefly overview the key steps just to help understand the main idea. For further details, readers are referred to \cite{Beuthe-2008:thin,Beuthe-2018:enceladuss} and \cite{Beuthe-2019:enceladuss}. The calculation has two steps: first calculating the tidal heating as if the ice shell had no spatial variation, and second adjusting the heating rate to account for the inhomogeneity.

On Enceladus, tidal forcing is mainly eccentricity-induced. As derived in \cite{Tyler-2011:tidal}, the tidal forcing consists of three spherical harmonic modes, $Y_{2,0},\ Y_{2,2}$ and $Y_{2,-2}$. Since the interaction between different tidal modes can only redistribute heating in zonal direction (the cross-interference terms only project onto $Y_{2,2},\ Y_{4,2}$ and $Y_{4,4}$) and our focus is latitudinal distribution of the longitudinally averaged tidal heating, we can calculate the tidal heating induced by the three tidal forcing modes separately and add them together. Also, because the planet is symmetric between east and west ignoring any asymmetry induced by the weak self-rotation, we can replace the $Y_{2,2}$ and $Y_{2,-2}$ modes with one equivalent $Y_{2,2}$ mode,
\begin{eqnarray}
  \label{eq:tidal-potential-combineY22}
  U_{\mathrm{eccen}}=\Re\left[\left(A_{2,0} Y_{2,0}(\theta,\phi)+ A_{2,2}Y_{2,2}(\theta,\phi)\right)e^{-i\omega t}\right].
\end{eqnarray}
Here, $\omega$ is the orbital angular velocity, and the amplitudes are given by
\begin{eqnarray}
    A_{2,0}= -\sqrt{\frac{4\pi}{5}} \frac{3}{2} a^2\omega^2 e,\  A_{2,2}=\sqrt{\frac{96\pi}{5}} \sqrt{\left(\frac{7}{8}\right)^2+\left(\frac{1}{8}\right)^2} a^2\omega^2 e,\nonumber
\end{eqnarray}
where $a,\ e$ are the radius and eccentricity of Enceladus.

The distortion of the ice shell by the tidal forcing can be estimated using the thin ice model \cite{Beuthe-2008:thin}\footnote{Unlike the original paper by Beuthe, we ignore the self attraction effect induced by the mass redistribution in the ocean. The geopotential induced by self attraction is inversely proportional to $(2l+1)$, where $l$ is the degree of the load, and thus can enhance large scale deformations in particular. Also, we ignore the geopotential anomaly induced by the surface deformation, $w$, which has been shown to be small constrained by a rigid ice shell above \cite{Beuthe-2016:crustal}. These effects turn out to only slightly affect the parameter regime for symmetry breaking, when the tidal heating generated in the non-uniform ice shell is parameterized by multiplying an ice thickness dependent factor to the heating profile for a uniform ice shell as \cite{Beuthe-2019:enceladuss}; but they would amplify large-scale tidal heating anomalies when the inihomogeneity of the ice shell is considered explicitly by solving Eq~(\ref{eq:thin-ice-w}) and Eq~(\ref{eq:thin-ice-w}) given a laterally varying $D_0$ and $\alpha_0$.}, which was derived from equilibrium of forces.

For an ice shell with a global uniform thickness $H_0$, one can solve a constant parameter PDE for the radial displacement $w$, and the auxiliary stress function $F$, given a tidal potential $U$.
\begin{eqnarray}
  D_0\Delta^{\prime 2} w-(1-\nu) D_0\Delta^{\prime }w+a^{3} \Delta^{\prime} F =a^{4} \rho_w U \label{eq:thin-ice-w}\\
  \alpha_0\Delta^{\prime 2}F-(1+\nu) \alpha_0\Delta^{\prime}F-\frac{1}{a} \Delta^{\prime} w=0,\label{eq:thin-ice-F}
\end{eqnarray}
where
\begin{eqnarray}
  \Delta^\prime&=&\Delta+2, \label{eq:def-Delta-operator}
\end{eqnarray}
and $\Delta$ is the spherical Laplacian $(\sin\theta)^{-1}\partial_\theta \sin\theta \partial_\theta + (\sin\theta)^{-2}\partial_\phi^2$. $\phi$ denotes longitude.
The definitions of extensibility $\alpha_0$ and the bending rigidity $D_0$ are in Table.~\ref{tab:parameters}.

Substituting $U_{\mathrm{eccen}}$ (Eq.~(\ref{eq:tidal-potential-combineY22})) into Eq.~(\ref{eq:thin-ice-w}) and Eq.~(\ref{eq:thin-ice-F}) yields,
\begin{eqnarray}
  F&=&\sum_{(l,m)\in \{(2,0),(2,2)\}}\tilde{F}^{(l)} A_{l,m}Y_{l,m} \label{eq:F0}\\
  w&=&\sum_{(l,m)\in \{(2,0),(2,2)\}}\tilde{w}^{(l)} A_{l,m} Y_{l,m},\label{eq:w0}
\end{eqnarray}
where
\begin{eqnarray}
\tilde{F}^{(l)}&=&\frac{a^3\rho_w }{\delta(l)} \\
\tilde{w}^{(l)}&=&\frac{\alpha_0(d(l)-1-\nu) a^4\rho_w }{\delta(l)}\\
  \delta(l)&=&a^2 d(l)+D_0\alpha_0 d(l)\left[(d(l)-1)^2-\nu^2\right]\label{eq:def-delta}\\
  d(l)&=&-l(l+1)+2=-(l-1)(l+2). \label{eq:def-d}
\end{eqnarray}

Following \cite{Beuthe-2019:enceladuss}, we first calculate the dissipative heating generated in the global homogeneous ice shell with thickness $H_0$. The heating includes three components: membrane mode $\mathcal{T}_0^{\mathrm{mem}}$, mixed mode $\mathcal{T}_0^{\mathrm{mix}}$ and bending mode $\mathcal{T}_0^{\mathrm{bend}}$).
\begin{eqnarray}
  \mathcal{T}^{\mathrm{mem}} &=&-\frac{\omega}{2} \mathrm{Im}(\alpha)\left(\left|\Delta^{\prime} F\right|^{2}-(1+\nu) \mathcal{A}\left(F ; F^{*}\right)\right) \label{eq:Heat-mem}\\
  \mathcal{T}^{\mathrm{mix}} &=&\frac{\omega}{2a} \mathrm{Im}(\chi)\left(\mathcal{A}\left(F ; w^{*}\right)+\mathcal{A}\left(F^{*} ; w\right)\right) \label{eq:Heat-mix} \\
  \mathcal{T}^{\mathrm{bend}} &=&\frac{\omega}{2a^4} \mathrm{Im}(D)\left(\left|\Delta^{\prime} w\right|^{2}-(1-\nu) \mathcal{A}\left(w ; w^{*}\right)\right). \label{eq:Heat-bend}
\end{eqnarray}
Here, $\mathcal{A}$ is a differential operator defined in \cite{Beuthe-2008:thin},
\begin{eqnarray}
  \mathcal{A}(a ; b)&=& \frac{1}{4}\left[-\Delta^{\prime} \Delta^{\prime}(a b)-\left(\Delta^{\prime} \Delta^{\prime} a\right) b-a\left(\Delta^{\prime} \Delta^{\prime} b\right)\right.\nonumber\\
                    &+&2\left(\Delta^{\prime} a\right)\left(\Delta^{\prime} b\right)+2 \Delta^{\prime}\left(\left(\Delta^{\prime} a\right) b+a\left(\Delta^{\prime} b\right)\right) \nonumber\\
                    &~&\left.-2\left(\Delta^{\prime}(a b)+\left(\Delta^{\prime} a\right) b+a\left(\Delta^{\prime} b\right)\right)+8 a b\right] \label{eq:def-A-operator}
\end{eqnarray}

Substituting Eq.~(\ref{eq:F0}) and Eq.~(\ref{eq:w0}) into the dissipative heating formula (Eq.\ref{eq:Heat-mem0}-\ref{eq:Heat-bend}), we get
\begin{eqnarray}
    \mathcal{T}^{\mathrm{mem}}_0 &=&-\frac{\omega}{2} \mathrm{Im}(\alpha_0) \sum_* A_{l,m}^2 \left|\tilde{F}_0^{(l)}\right|^{2}   \left[d(l)^2E(l,m,l,-m,j)-(1+\nu) A(l,m,l,-m,j)\right] Y_{j,0} \label{eq:Heat-mem0}\\
  \mathcal{T}^{\mathrm{mix}}_0 &=&\frac{\omega}{a} \mathrm{Im}(\chi_{0})\sum_* A_{l,m}^2 \mathrm{Re}\left(\tilde{F}_0^{(l)}\tilde{w}_0^{(l)*}\right)  A(l,m,l,-m,j)Y_{j,0} \label{eq:Heat-mix0} \\
  \mathcal{T}^{\mathrm{bend}}_0 &=&\frac{\omega}{2a^4} \mathrm{Im}(D_0) \sum_*  A_{l,m}^2 \left|\tilde{w}_0^{(l)}\right|^{2} \left[d(l)^2E(l,m,l,-m,j)-(1-\nu) A(l,m,l,-m,j)\right] Y_{j,0},  \label{eq:Heat-bend0}
\end{eqnarray}
where
\begin{eqnarray}
    &~&A(l_1,m_1,l_2,m_2;j)\equiv\langle \mathcal{A}(Y_{l_1,m_1};Y_{l_2,m_2})\cdot Y^*_{j,m_1+m_2}\rangle\label{eq:def-A}\\
  &~&E(l_1,m_1,l_2,m_2;j)\equiv \langle \left(Y_{l_1,m_1}Y_{l_2,m_2}\right)\cdot Y^*_{j,m_1+m_2}\rangle\label{eq:def-E}\\
  &~&\sum_* \equiv \sum_{(l,m)\in \{(2,0),(2,2)\}}\sum_{0\leq j \leq 2l}\nonumber
\end{eqnarray}
The total tidal heating is the sum over the three modes
\begin{equation}
  \label{eq:Heat-total0}
  \mathcal{T}_0=\mathcal{T}^{\mathrm{mem}}_0+\mathcal{T}^{\mathrm{mix}}_0+\mathcal{T}^{\mathrm{bend}}_0.
\end{equation}

According to Eq.~(\ref{eq:ice-evolution}), the time scale for tidal heating to change ice topography is
\begin{equation}
  \label{eq:tau-T}
  \tau_{\mathcal{T}}\equiv \left(\frac{d(H/H_0)}{dt}\right)_{\mathcal{T}}^{-1}\sim \frac{L_f\rho_iH_0}{\gamma\overline{\mathcal{T}_0}},
\end{equation}
where $\overline{(\cdot)}$ denotes the global average.

The tidal heating generated in a laterally varying ice shell is then approximated by multiplying an ice thickness dependent factor to the membrane mode as in \cite{Beuthe-2019:enceladuss}. 
\begin{eqnarray}
  \label{eq:Heat-nonuniform-ours}
  \mathcal{T} \approx \left(\frac{H}{H_0}\right)^{p_\alpha} \mathcal{T}^{\mathrm {mem }}_0+\mathcal{T}^{\mathrm {mix }}_0+\mathcal{T}^{\mathrm {bend }}_0
\end{eqnarray}
Here, $p_\alpha$ is a tunable parameter. In \cite{Beuthe-2019:enceladuss}, $p_\alpha$ is set to $-1$, because extensibility $\alpha\sim H^{-1}$ and tidal dissipation is proportional to $\alpha$ times the square of tidal load. We can get some physical intuition by considering a rubber band with a weak point on it. When stretched, the weak point will deform more and thus generate more heat. 

Besides geometrical factors, regions with thinner ice will also be warmer and thus less rigid. That will help generate more heat in regions where it is already warm (rheology feedback). In \cite{Beuthe-2019:enceladuss}, this effect is explicitly considered by iterating the tidal heating and the temperature profile of the ice shell.
In Beuthe's calculation, the ice shell over the north pole is about half as thick as the ice shell over the equator, and that would double the heat generation over the north pole relative to the equator on top of polar-amplified heat profile obtained assuming uniform ice thickness. The rheology feedback was found to be as high as 60\% at the north pole (even more for the south pole) and less than 10\% at the equator\footnote{We refer to the ISO-H case in \cite{Beuthe-2019:enceladuss}, where the author amplify the ice shell dissipation by a factor of 10, and we do the same here.}. Therefore, geometric effects and rheology feedback would enhance the shell dissipation at the north pole $2\times(1+60\%)/(1+10\%)\sim 3$ times over that at the equator, and this leads to a $p_\alpha\sim -\log_2(3)=-1.6$.

In addition, a more dissipative ice shell would have an even stronger rheology feedback \cite{Beuthe-2019:enceladuss}, and ocean circulation may also contribute. We anticipate that the relatively low (high) salinity induced by melting (freezing) may help drive an ocean circulation that is upwelling (sinking) in the melting (freezing) zones, further amplifying the melting and freezing. Given these uncertainties, we leave $p_\alpha$ as a tunable parameter that is more negative than $-1$.

\begin{figure}
  \centering
  \includegraphics[width=0.5\textwidth]{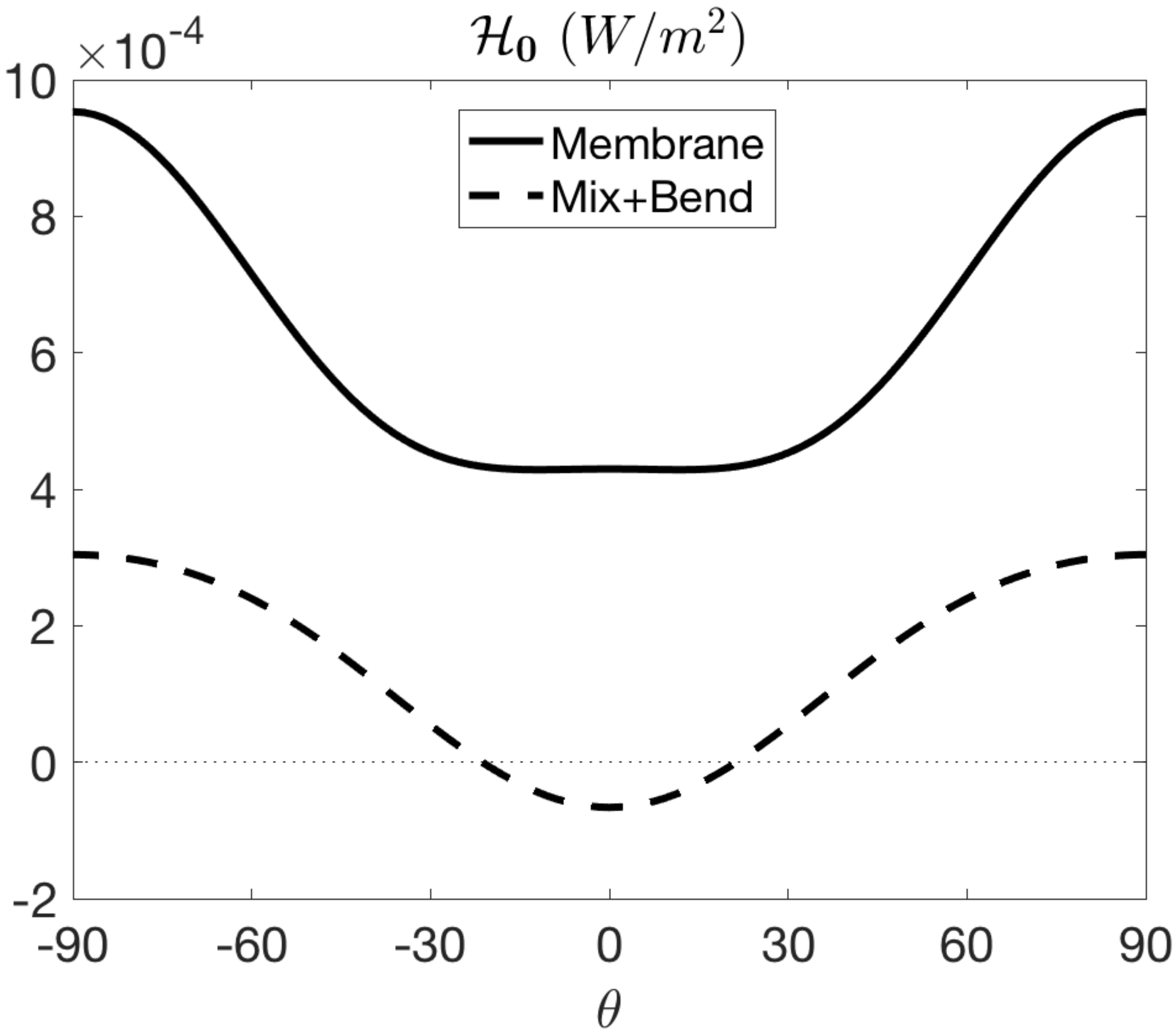}
  \caption{The meridional profile of the tidal heating generated in a uniform ice shell. }
  \label{fig:H0}
\end{figure}

\subsection*{Linearized equation for ice thickness evolution and the unstable mode}

Perturbing Eq.~(\ref{eq:ice-evolution}) around the equilibrium profile $h_{eq}'$ and dropping $\mathcal{B}$ and $\mathcal{C}$, we get
 \begin{eqnarray}
   \frac{dh''}{dt}&=&-\frac{\mathcal{F}_0}{L_fH_0\rho_i}(1+h_{eq}')^{-2}h'' -p_\alpha\frac{\gamma\mathcal{T}_0^{\mathrm{mem}}}{L_fH_0\rho_i}(1+h_{eq}')^{p_\alpha-1}h''\nonumber\\
                  &+&\mathcal{Q}_0\left.\left[\left(\partial_\theta^2(1+h_{eq}')^3\right) h'' +2\left(\partial_\theta(1+h_{eq}')^3\right)\left(\partial_\theta h''\right) + (1+h_{eq}')^3 \partial_\theta^2h'' \right]\right/a^2\nonumber\\
                  &+&\mathcal{Q}_0\frac{\cot\theta}{a^2}\left[\left(\partial_\theta(1+h_{eq}')^3\right) h'' + (1+h_{eq}')^3 \partial_\theta h'' \right].\nonumber\\
                  &=&\left[-\frac{\mathcal{F}_0}{L_fH_0\rho_i}(1+h_{eq}')^{-2} - p_\alpha\frac{\gamma\mathcal{T}_0^{\mathrm{mem}}}{L_fH_0\rho_i}(1+h_{eq}')^{p_\alpha-1}+\frac{\mathcal{Q}_0}{a^2}\left(\partial_\theta^2(1+h_{eq}')^3\right) \right]h''\nonumber\\
                  &+&\frac{\mathcal{Q}_0}{a^2}\left[ 2\left(\partial_\theta(1+h_{eq}')^3\right)\left(\partial_\theta h''\right) + (1+h_{eq}')^3 \partial_\theta^2h'' \right]\nonumber\\
                  &+&\frac{\mathcal{Q}_0}{a^2}\cot\theta \left[\left(\partial_\theta(1+h_{eq}')^3\right) h'' + (1+h_{eq}')^3 \partial_\theta h'' \right].\label{eq:linearized-ice-evolution}
 \end{eqnarray}
 The above linearized dynamic system can be written in a matrix form through finite differences.
 The crack-induced cooling $\mathcal{C}$ can be ignored because the ice shell thickness in equilibrium state is greater than a threshold $H_{\mathrm{crack}}=8$~km everywhere. 
 The balancing term $\mathcal{B}$ can be accounted by multiplying a matrix $\left[\mathbf{I}-\mathbf{1}\mathbf{w}\right]$ to the matrix form corresponding to the above linear system $\mathbf{M}$, where $\mathbf{w}=\mathrm{diag}(\sin\theta/N_\theta)$ is the area weight matrix with each grid cell's weight written on the diagonal, $\mathbf{I}$ is the identity matrix, and $\mathbf{1}$ is the all-one matrix.
 
The most unstable eigenmode (shown in Fig.~2b of the main text) has a pole-to-pole tilting structure because this structure is least damped by the ice flow, and meanwhile, is most strongly enhanced by the tidal heating. This can be seen more clearly under a simpler configuration. We drop the spherical curvature terms, and ignore the inhomogeneity of the ice thickness in the equilibrium state. We also assume that the bending mode and mixed mode are negligible, and simplify the $\mathcal{T}_0^{\mathrm{mem}}$ profile by dropping the relatively small $\cos(4\theta)$ component. These simplifications lead to a tangential linear system for the perturbation  $h''$ on top of the equilibrium state,
\begin{eqnarray}
  \frac{dh''}{dt} &=&-\left[ \frac{ \gamma p_\alpha(\tilde{\mathcal{T}}_0^{\mathrm{mem, k=2}}+\mathcal{T}_0^{\mathrm{mem, k=0}})+\mathcal{F}_0}{L_fH_0\rho_i}\right] h''+\frac{\mathcal{Q}_0}{a^2} \partial_\theta^2h'', \label{eq:simple-linearized-ice-evolution}
 \end{eqnarray}
 where $\tilde{\mathcal{T}}_0^{\mathrm{mem, k=2}}$ and $\tilde{\mathcal{T}}_0^{\mathrm{mem, k=0}}$ denotes the amplitude of the wavenumber 2 (with a structure that peaks at the two poles) and wavenumber 0 (constant over the globe) Fourier component of $\mathcal{T}_0^{\mathrm{mem}}$. Expanding the perturbation $h''$ into a cosine series\footnote{Using cosine series guarantees the boundary condition: $\partial_\theta h''$, and hence ice flow, vanishes at the two poles.} yields
\begin{eqnarray}
  \sum_k \frac{d h_k''}{dt}\cos(k\theta)&=&-\sum_k \frac{\gamma p_\alpha\tilde{\mathcal{T}}_0^{\mathrm{mem}, k=2}}{L_fH_0\rho_i}\cos(2\theta)\cdot h_k''\cos(k\theta) \nonumber\\
  &~&- \sum_k \frac{\gamma p_\alpha\tilde{\mathcal{T}}_0^{\mathrm{mem}, k=0}+\mathcal{F}_0}{L_fH_0\rho_i}\cdot h_k''\cos(k\theta) -\sum_k \frac{\mathcal{Q}_0k^2}{a^2}h_k''\cos(k\theta) \nonumber\\
  \frac{d h_k''}{dt}&=& -\frac{\gamma p_\alpha\tilde{\mathcal{T}}_0^{\mathrm{mem}, k=2}}{2L_fH_0\rho_i}(h_{k+2}''+h_{|k-2|}'')- \frac{\gamma p_\alpha\tilde{\mathcal{T}}_0^{\mathrm{mem}, k=0}+\mathcal{F}_0}{L_fH_0\rho_i} h_k''-\frac{\mathcal{Q}_0k^2}{a^2}h_k''. \label{eq:symmetry-breaking-mechanism}
\end{eqnarray}
The ice flow term (the third term) damps perturbations with all wavenumbers, but the gravest mode ($k=1$) is least damped. Meanwhile, the perturbations grow because of tidal heating (the first term, tidal heating minus heat loss). Since the tidal heating in a homogeneous ice shell, $\mathcal{T}_0^{\mathrm{mem}}$, peaks at the poles (i.e., dominated by  the $\cos(2\theta)$ component), the wavenumber of perturbation being amplified is offset by a wavenumber $2$ from that of the existing perturbation. Therefore, a perturbation with a specific wavenumber can only grow through its interaction with other wavenumbers. $k=1$ is the only exception because $|k-2|$ equals $1$; this makes $k=1$ the fastest growing mode.

Symmetry breaking requires the ice flow to be slow enough to allow the $k=1$ mode to grow, and also requires the ice flow to be fast enough that the equilibrium ice topography $H_{\mathrm{eq}}$ forced by the polar-amplified tidal heating profile (corresponding to the $k=2$ mode) is crack-free (once cracks form, the strong crack-induced cooling will fix the ice thickness to $H_{\mathrm{crack}}$).
Ignoring the mode interactions, the criteria for symmetry breaking can be written as
\begin{eqnarray}
  -\frac{\gamma p_\alpha\tilde{\mathcal{T}}_0^{\mathrm{mem}, k=2}}{2L_fH_0\rho_i}- \frac{\gamma p_\alpha\tilde{\mathcal{T}}_0^{\mathrm{mem}, k=0}+\mathcal{F}_0}{L_fH_0\rho_i}-\frac{\mathcal{Q}_0}{a^2}&>&0 \hspace{2cm} \mathrm{(k=1~grow)}\\
   \frac{\gamma\tilde{\mathcal{T}}_0^{\mathrm{mem}, k=2}}{L_fH_0\rho_i}-\frac{4\mathcal{Q}_0}{a^2}&<&0 \hspace{2cm} (H_{eq}~\mathrm{is~crack-free}).
\end{eqnarray}
Let $\beta= \mathcal{T}_0^{\mathrm{mem}, k=2}/\mathcal{T}_0^{\mathrm{mem}, k=0}$ and substitute $\gamma\mathcal{T}_0^{\mathrm{mem}, k=0}/\mathcal{F}_0= \tau_{\mathcal{F}}/\tau_{\mathcal{T}}$\footnote{Note that $\mathcal{T}_0^{\mathrm{mem}}$ should be positive everywhere and that the tidal heating generated in the ice shell $\mathcal{T}_0^{\mathrm{mem}, k=0}$ cannot exceed the heat loss rate $\mathcal{F}_0$. Both $\beta$ and $\tau_{\mathcal{F}}/\tau_{\mathcal{T}}$ should be between 0 and 1.}, the above condition can be rewritten as
\begin{eqnarray}
  \label{eq:symmetry-breaking-criteria}
  \frac{\beta(\tau_{\mathcal{F}}/\tau_{\mathcal{T}})}{4}<\frac{\tau_{\mathcal{F}}}{\tau_{\mathcal{Q}}}<-p_\alpha(\tau_{\mathcal{F}}/\tau_{\mathcal{T}})(\beta/2+1)-1
\end{eqnarray}
The definitions of $\tau_{\mathcal{F}}$ and $\tau_{\mathcal{Q}}$ are given in Eq.~(\ref{eq:tau-F}) and Eq.~(\ref{eq:tau-Q}). As shown in Fig.~\ref{fig:H0}, the variation of $\mathcal{T}_0^{\mathrm{mem}}$ across different latitudes is comparable to its mean value, so $\beta\sim 1$. For such a $\frac{\tau_{\mathcal{F}}}{\tau_{\mathcal{Q}}}$ to exist, the upper bound of the above inequality has to be larger than the lower bound,
\begin{equation}
  \label{eq:symmetry-breaking-criteria2}
  \frac{\tau_{\mathcal{F}}}{\tau_{\mathcal{T}}}>\frac{1}{-p_\alpha(\beta/2+1)-\beta/4}.
\end{equation}
This is consistent with physical intuition: conduction always tends to damp the ice topography, so topographic growth is only possible when the tidal heating generated in the ice shell contributes a significant part in balancing the heat loss to space.
 
Eq.~\ref{eq:symmetry-breaking-criteria} and Eq.~\ref{eq:symmetry-breaking-criteria2} together determine the location of the symmetry breaking regime in the parameter phase space.


\bibliography{export}
 \bibliographystyle{Science}